\documentclass[aps,prb,twocolumn,showpacs]{revtex4-1}

\usepackage{amssymb}
\usepackage{amsmath}
\usepackage{amsthm}

\usepackage{graphicx}
\usepackage{epstopdf}
\usepackage{xspace}
\usepackage{ulem}

\graphicspath{{./figures/}}

\begin{document}
\title{The influence of material defects on current-driven vortex domain wall mobility}

\author{Jonathan Leliaert$^{1,2}$, Ben Van de Wiele$^1$, Arne Vansteenkiste$^2$, Lasse Laurson$^3$, Gianfranco Durin$^{4,5}$, Luc Dupr\'e$^1$, Bartel Van Waeyenberge$^2$}
\affiliation{$^1$Department of Electrical Energy, Systems and Automation, Ghent University, Ghent B-9000, Belgium}
\affiliation{$^2$Department of Solid State Science, Ghent University, Krijgslaan 281/S1 , 9000 Ghent, Belgium.}
\affiliation{$^3$COMP Centre of Excellence and Helsinki Institute of Physics, Department of Applied Physics, Aalto University School of Science, P.O. Box 11100, FI-00076 AALTO, Finland}
\affiliation{$^4$Istituto Nazionale di Ricerca Metrologica, Strada delle Cacce 91, 10135 Torino, Italy.}
\affiliation{$^5$ISI Foundation, Via Alassio 11/c, 10126, Torino, Italy.}

\begin{abstract}
Many future concepts for spintronic devices are based on the current-driven motion of magnetic domain walls through nanowires.   Consequently a thorough understanding of the domain wall mobility is required.  However, the magnitude of the non-adiabatic component of the spin-transfer torque driving the domain wall is still debated today as various experimental methods give rise to a large range of values for the degree of non-adiabaticity.  Strikingly, experiments based on vortex domain wall motion in magnetic nanowires consistently result in lower values compared to other methods.  Based on the micromagnetic simulations presented in this contribution we can attribute this discrepancy to the influence of distributed disorder which vastly affects the vortex domain wall mobility, but is most often not taken into account in the models adopted to extract the degree of non-adiabaticity.  
\end{abstract}

\maketitle

\section{INTRODUCTION}
In many future spintronic devices information is stored and processed by means of magnetic domain walls moving through magnetic nanowires \cite{ALL-05,BAR-06,PAR-08}.  Here, electrical currents are able to drive the magnetic domain walls by means of the spin-transfer torque interaction.  To correctly describe this interaction, next to an adiabatic term also a non-adiabatic term \cite{ZHA-04}, should be added to the Landau-Lifshitz equation,
\[
\begin{split}
\mathbf{\dot{m}}=&\gamma\mathbf{H_\mathrm{eff}}\times\mathbf{m}+\alpha \mathbf{m}\times\mathbf{\dot{m}}\\
&-\left[\mathbf{u}\cdot\nabla\right]\mathbf{m}+\beta \mathbf{m}\times \left[\mathbf{u}\cdot\nabla\right]\mathbf{m}.
\end{split}
\]
Here $\mathbf{m}$ is the magnetisation, $\gamma$ the gyromagnetic ratio, $\alpha$ the Gilbert damping constant, $\mathbf{u}$ a velocity-like term proportional to the current density $J$ and $\beta$ the degree of non-adiabaticity.  Since the introduction of this non-adiabatic term, there has been a lot of debate on the magnitude of $\beta$, with theoretically predicted values ranging from $\beta\approx\alpha$\cite{TSE-06,ZHA-04,BER-08} over $\beta=2\alpha$ \cite{THI-05} to $\beta=4\alpha$ \cite{BAN-09}. Additionally, experiments have been until now unable to converge to one value. \\

Several experimental techniques have been used to quantify $\beta$.  One way is to measure the depinning field to pull a vortex out of a pinning potential in the presence of a spin-polarized current \cite{LEP-09,LEP-09a,LEP-10}. A similar technique consists of looking at the thermal hopping between pinning sites in the presence of a spin-polarized current \cite{ELT-10} where different values for $\beta$ for the same material are estimated depending on the considered magnetic structure: a vortex domain wall or a transverse domain wall. Another approach is to determine local vortex core displacements due to spin-polarized currents in the confining potential of e.g. a pinning site \cite{THO-06}, a disk \cite{HEY-10a}, or a square geometry \cite{POL-12}.  A third set of experiments, only able to extract $\beta/\alpha$, is based on measuring the distance a domain wall is able to cover due to a current pulse with known amplitude and duration.  Here, resulting time and space averaged velocities are fitted to theoretical and/or simulated values \cite{HEY-10,BEA-06,HAY-06,MEI-07}. Apart from these methods to directly quantify $\beta$ or $\beta/\alpha$, electrical and magnetic imaging techniques show domain wall transformations when an electric current is applied, indicating $\beta\neq\alpha$ \cite{BOU-11,KLA-05,HEY-08}. Table \ref{tabel} gives an overview of experiments performed to measure $\beta$.

\renewcommand{\thefootnote}{\ifcase\value{footnote}\or(*)\or(**)\or(***)\or(****)\or(\#)\or(\#\#)\or(\#\#\#)\or(\#\#\#\#)\or($\infty$)\fi}
\begin{table}[!htb]
\caption{\label{tabel} Overview of experimentally obtained values for $\beta$ in Permalloy.}
\begin{tabular}{l|l|l|l}
Method&$\beta$&$\beta/\alpha$&Ref.\\
\hline
Current-assisted domain &$0.040\pm0.005$ &2*&\cite{LEP-09}\\
wall depinning from a &$0.040\pm0.005$ &2-4*&\cite{LEP-09a}\\
pinning site &$0.040\pm0.0025$ &$\approx5.3$&\cite{LEP-10}\\
\cline{1-4}
Thermal depinning& & &\\
\hspace{1.5mm}vortex domain wall & $0.073\pm 0.026$ & $\approx 9$ & \cite{ELT-10}\\
\hspace{1.5mm}transverse domain wall & $0.01\pm 0.004 $&$\approx 1$ &\cite{ELT-10}\\\cline{1-4}
Local vortex core& $0.04$ &8*&\cite{THO-06}\\
 movements & $0.15 \pm0.07$ &$>$10&\cite{HEY-10a}\\
& $0.15 \pm0.02$ &$>7$&\cite{POL-12}\\
\hline
Vortex domain wall &0.01* &$0.96\pm0.02$&\cite{MEI-07}\\
motion in&0.008* &1&\cite{HAY-06}\\
nanowires& 0.007*&0.7&\cite{BEA-06}\\
&not mentioned&1&\cite{HEY-10}\\
\multicolumn{3}{l}{* based on estimated values of $\alpha$}\\
\end{tabular}
\end{table}
Even within the broad range of possible values reported, a clear discrepancy between measurements based on domain wall motion and other methods is present. Here we show that these seemingly discrepant values for $\beta$ can be ascribed to the influence of distributed disorder on the time and space averaged motion of the magnetic domain wall, giving rise to an apparent degree of non-adiabaticity $\beta \approx\alpha$ irrespective of the actual value of $\beta$. This is supported by simulations of the motion of vortex domain walls in nanowires including the effect of realistic distributed disorder. \\

Simulations investigating the effect of sample imperfections on the domain wall mobility have mainly concentrated on nanowire edge roughness \cite{NAK-03,MAR-12a}. It is found that this suppresses the Walker breakdown (defined by the maximal linear motion of the transverse domain wall), allowing the domain wall to move faster for higher applied fields or currents compared to the corresponding nanowire with perfect geometry. 
These studies however neglect the influence of disorder distributed within the wire. Nevertheless, real Permalloy nanowire samples contain defects in their microstructure, e.g., surface roughness and/or grain boundaries, which can act as pinning centres for the domain walls. From experiments \cite{COM-06,COM-10,KIM-10b,BUR-13,CHE-12} it is known that these are randomly distributed throughout the wire with a density $\sigma$ ranging from 690 to 2000 $\mu m^{-2}$, and give rise to a pinning potential for a vortex that is approximately 2 eV deep and has an interaction range roughly equal to the size of the vortex core. In this contribution we numerically investigate the influence of such distributed disorder on the domain wall mobility.

\begin{figure}
\centering
\includegraphics[width=1\linewidth]{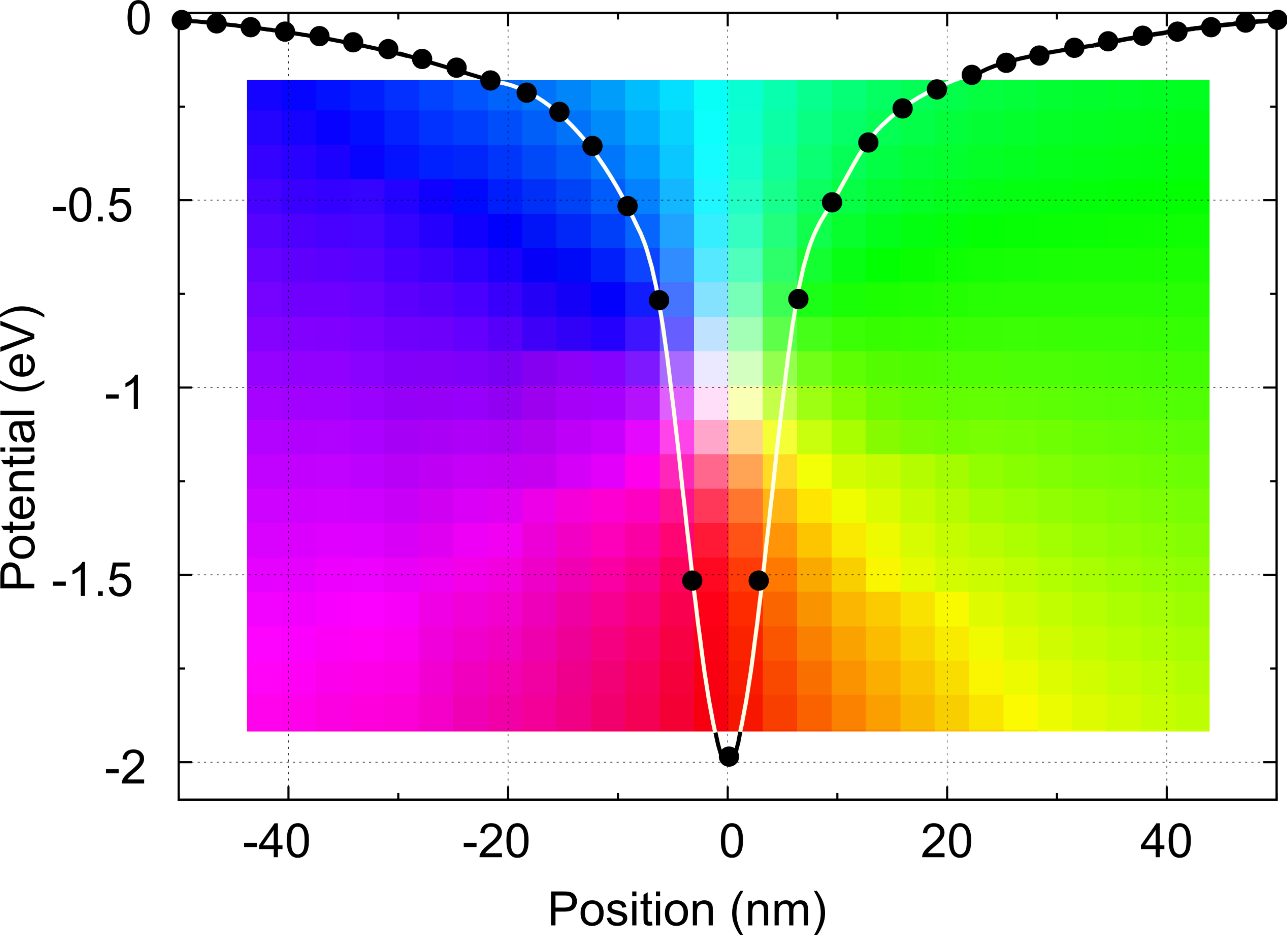}
\caption{\label{Fig_potential} The potential well of a defect of size $3\times\,3$ discretization cells interacting with a vortex core, with the exchange length reduced by roughly 50\% at the boundaries. The depth of the potential is approximately 2\,eV and the interaction range is comparable to the size of the vortex core diameter\cite{ik} .}
\end{figure}

\section{MICROMAGNETIC METHODS}
The micromagnetic simulations are performed using the software package \cite{VAN-11} \textsc{MuMax$^3$}. The domain wall motion is simulated in nanowires of width 400\,nm and thickness 10\,nm for 500\,ns, corresponding to a maximum wire length of 500 $\mu$m.  A discretization of $3.125\times3.125\times10$ nm$^3$ is used. The window in which the magnetisation is calculated is 1200\,nm wide and moves with the domain wall. Additionally, typical material parameters for Permalloy are used: saturation magnetisation $M_s=860 \times 10^3$\,A/m, Gilbert damping $\alpha=0.02$ and exchange stiffness $A = 13 \times 10^{-12}$\,J/m.  To see the influence of $\beta$ on the domain wall dynamics, values $\beta=0$, $\beta=\alpha$ and $\beta=2\alpha$ are considered. Here, the magnetic charges at the ends of the nanowire are compensated to simulate an infinitely long wire. For different current densities, the average domain wall velocity is computed based on the distance travelled by the wall during the simulation time. 
To include distributed defects in our simulations we introduced small regions ($9.375 \times\,9.375$ nm$^2$ in size) with a reduced exchange length 
\[
	l_{ex}=\sqrt{\frac{2 A}{\mu_0 M_s^2}}
\]
 at their boundaries. By reducing the exchange constant A to 30\% of its normal value across the region boundaries, a corresponding reduction in $l_{ex}$ of roughly 50\% was obtained. This method is a realistic way to include distributed defects that are reminiscent of material grains \cite{ik}.

In Fig. \ref{Fig_potential} the pinning potential for such a single region is shown, illustrating the correspondence with experimental values. Random distributions of these regions with densities ranging from $\sigma\,=\,500$ to $1500\,\mu m^{-2}$ were included. This differs from earlier approaches to simulate distributed disorder \cite{VAN-12, MIN-10}.
In Ref. \cite{VAN-12} disorder was introduced in the material by introducing void cells. Ref. \cite{MIN-10}, on the other hand, implemented disorder by introducing slight variations in the saturation magnetisation.

\section{RESULTS AND DISCUSSION}
\begin{figure}
\centering
\includegraphics[width=1\linewidth]{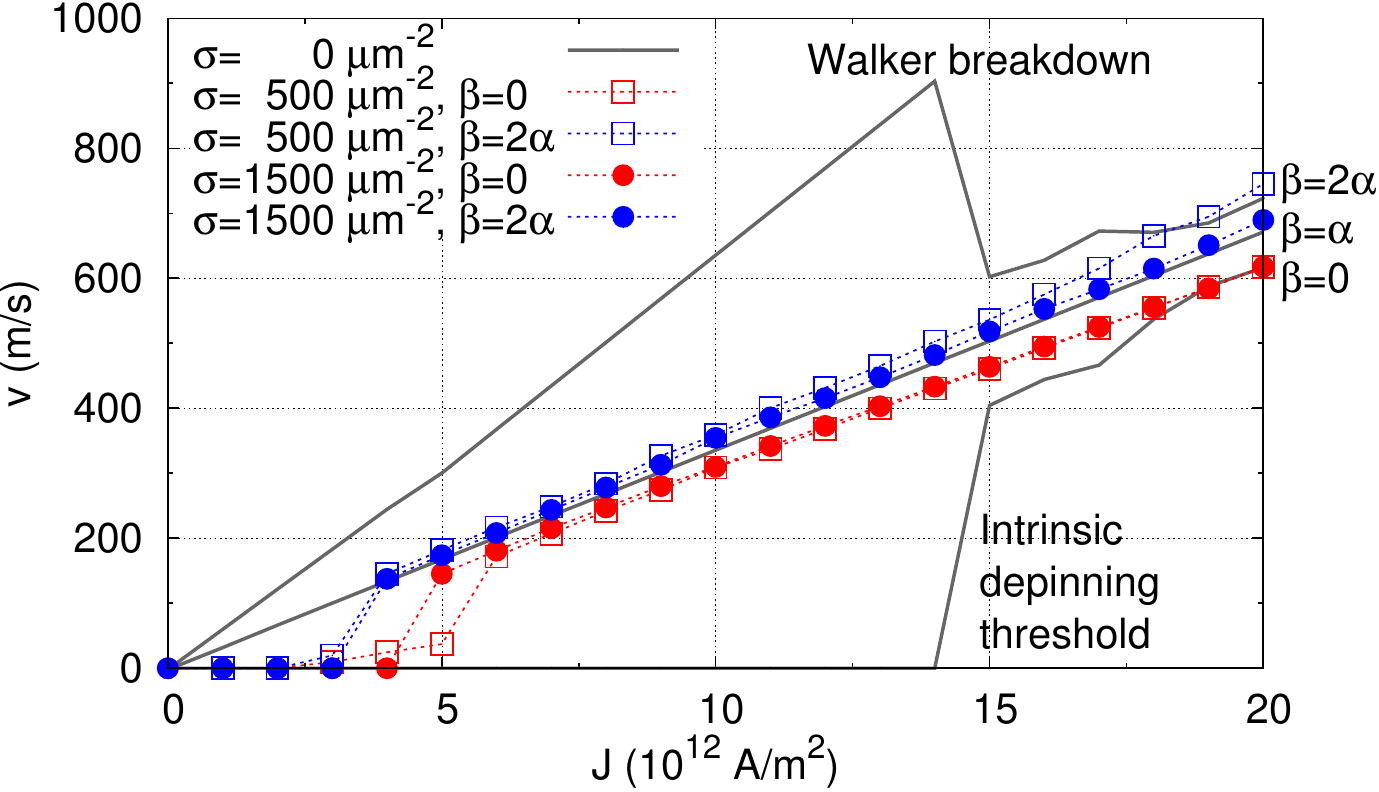}
\caption{\label{Fig_disorder} 
Velocity vs. applied current density. Solid lines: Velocity vs. applied current density in a perfect nanowire ($\beta=0$, $\alpha$ and $2\alpha$). Colored symbols: velocity vs. applied current density in nanowires with disorder for $\beta=0$ (red) and $\beta=2\alpha$ (blue).  Irrespective of the used value of $\beta$, the velocity curves tend to converge to a case corresponding to $\beta=\alpha$ in perfect wires. For small applied current densities, extrinsic pinning of the vortex core on a defect takes place.}
\end{figure}

\begin{figure*}
\centering
\includegraphics[width=0.8\linewidth]{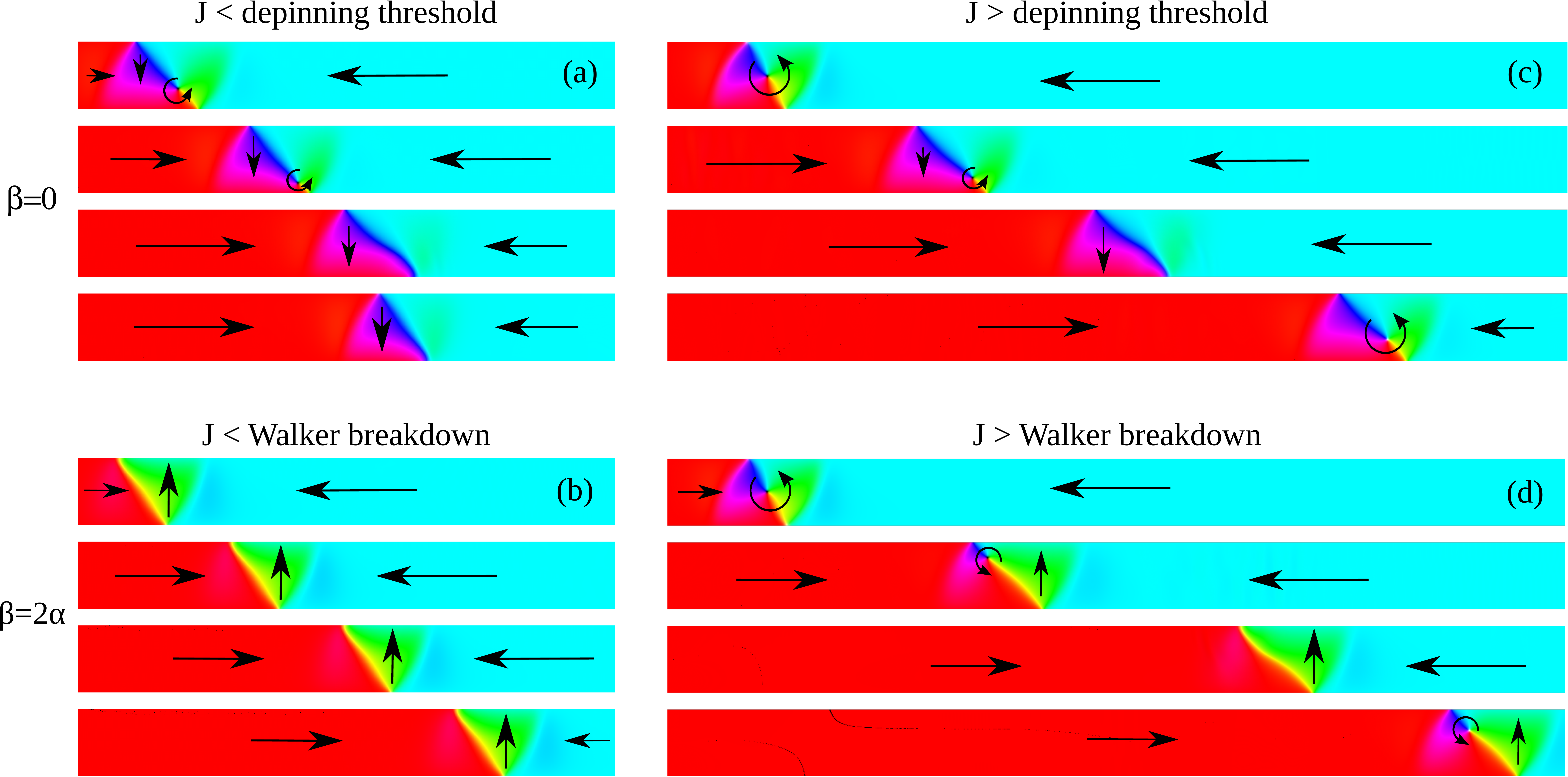}
\caption{\label{Fig_perfect} Snapshots of vortex domain wall motion in a perfect Py nanowire, 400\,nm wide and 10\,nm thick. While propagating, the vortex core moves towards the nanowire edge and the domain wall reshapes into a transverse domain wall. For small currents this wall gets intrinsically pinned, $\beta=0$ (a), or moves linearly with a speed proportional to applied current density, $\beta=2\alpha$ (b).  For large applied currents, the wall transforms again into a vortex domain wall, $\beta=0$ (c) and $\beta=2\alpha$ (d). Now the core has an opposite polarization and moves consequently towards the opposite nanowire edge giving rise to successive domain wall transformations.}
\end{figure*}

Figure \ref{Fig_disorder} shows averaged domain wall velocities $v$ versus applied current density $J$. Comparing the domain wall velocity in the disordered nanowires to the perfect nanowire case (solid lines) we observe (i)  a depinning threshold at much smaller currents than the intrinsic depinning threshold, for the adiabatic as well as for the non-adiabatic case and (ii) an absence of the Walker breakdown for the non-adiabatic case (here $\beta=2\alpha$). To find the origin of this very different behavior we compare the domain wall motion in a perfect wire and in a nanowire with distributed disorder, see Figs. \ref{Fig_perfect} and \ref{Fig_core_switches}. 

\subsubsection{PERFECT NANOWIRES}
 In a perfect wire, the vortex core moves towards the nanowire edge in the direction defined by the vortex core polarization (for $\beta\neq\alpha$). Below the intrinsic depinning threshold (for $\beta=0$) or the Walker breakdown (for $\beta>0$) the vortex domain wall reshapes into a transverse domain wall which gets intrinsically pinned or moves linearly with applied current density, respectively (see Supplementary Movies M1 and M2). Above the Walker breakdown/depinning threshold the vortex core switches polarization and moves to the opposite nanowire edge (see Supplementary Movies M3 and M4). For $\beta=\alpha$ the vortex core moves perfectly along the nanowire center.

\subsubsection{DISORDERED NANOWIRES}
\begin{figure}
\centering
\includegraphics[width=\linewidth]{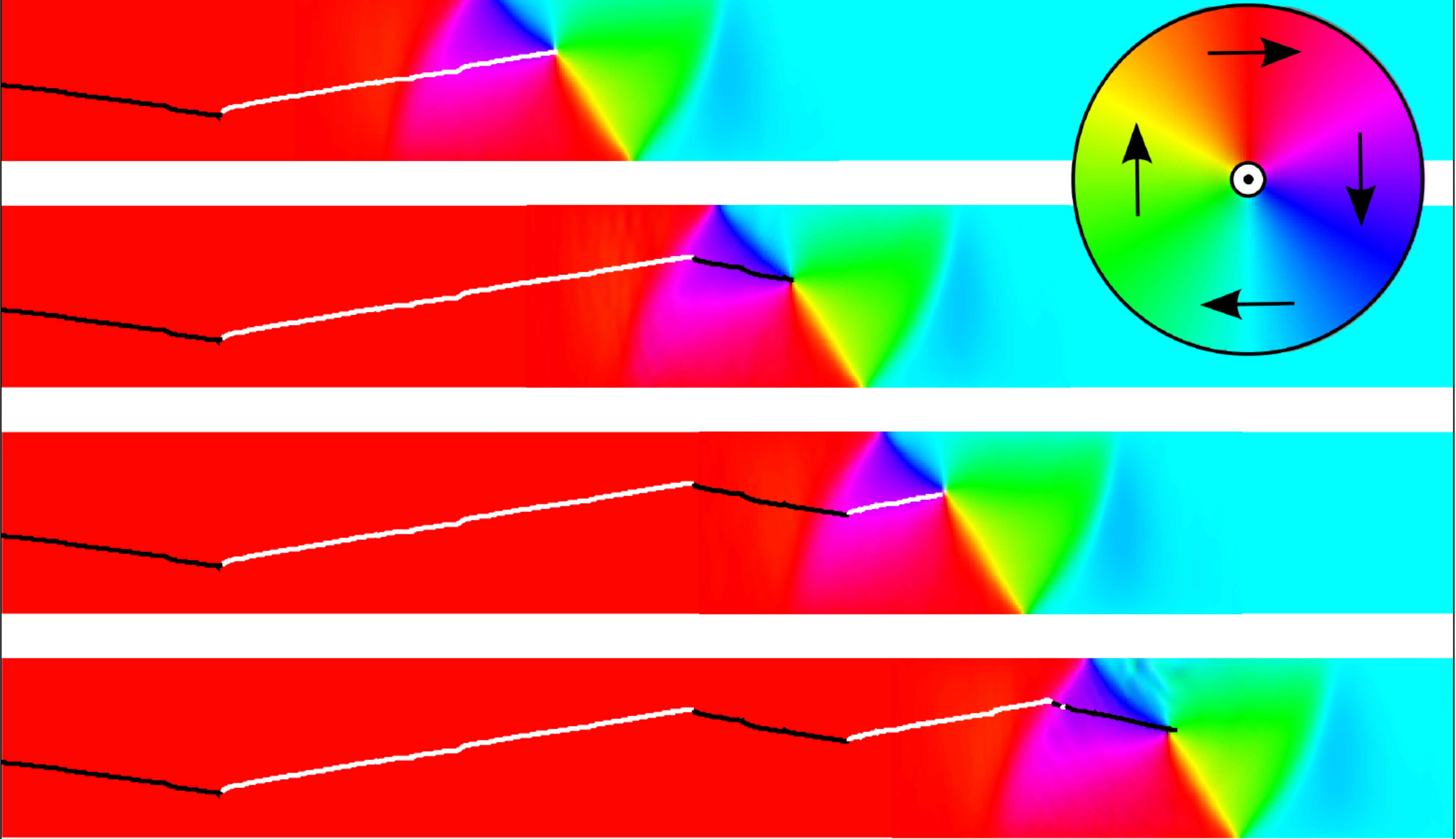}
\caption{\label{Fig_core_switches} 
Snapshots of vortex domain wall motion in a disordered Py nanowire, 400\,nm wide and 10\,nm thick. The successive magnetisation snapshots, at 3\,ns time intervals, show a vortex wall driven adiabatically ($\beta=0$) with current density $J=10\times10^{12}\,A/m^2$ in a nanowire with disorder density $\sigma=500\,\mu m^{-2}$. The vortex core trajectory is represented by the white/black lines, indicating a positive/negative vortex core polarization.}
\end{figure}

In a disordered nanowire (see Fig. \ref{Fig_core_switches} and Supplementary Movies M5 and M6), the vortex core can switch polarization at a defect, implying a change in lateral propagation direction and thus hindering the formation of the transverse domain wall. This polarization switching mechanism, which was not found \cite{STILES} in Ref. \cite{MIN-10}, explains the absence of the Walker breakdown and the much smaller depinning threshold. The pinning mechanism itself is also affected by the disorder: instead of the intrinsic pinning mechanism induced by the internal balancing of the effective field and spin-transfer torques inside the transverse domain wall found in a perfect nanowire, disorder gives rise to an extrinsic pinning mechanism in which the vortex core gets pinned at a defect. In the experimentally accessible current ranges, we observe an average motion of the vortex core in the central region of the wire without the formation of transverse domain walls due to successive core switches at defects. This resembles the motion of a vortex domain wall in a perfect wire for the case $\beta=\alpha$, which explains the values of $\beta/\alpha$ derived from domain wall motion (table \ref{tabel}).\\

Following the work of Ref. \cite{VAN-12} we have also investigated the effect of voids on the domain wall mobility and have found qualitatively the same mobility curves. This can be explained by the fact that the pinning potentials caused by this type of disorder are much deeper \cite{ik}, and therefore also allow vortex core polarization switching, resulting in the same type of motion. Thus, we expect that all types of defects that give rise to pinning potentials that are sufficiently deep (e.g. voids, grain boundaries, thickness fluctuations,\dots) allow vortex core polarization switching and consequently lead to the same mobility.\\

Due to the consecutive polarization switches at defects, the vortex core generally does not reach the edges of the nanowire. However, in the event it does happen, we observe that defects at the edges allow the nucleation of a vortex core of opposite polarity, as is the case in wires with edge roughness \cite{NAK-03}.\\

In contrast to the 0\,K temperature we considered in the simulations, a non-zero temperature results in thermally activated depinning and finite but small velocities in a creep regime \cite{MET-07}. We checked the influence of temperature on the observed phenomena. However, apart from introducing non-zero velocities slightly below the depinning threshold (creep regime), no influence was seen on the domain wall mobility in the flow regime.\\

\section{CONCLUSION}
In conclusion, we have shown that material defects vastly influence the domain wall dynamics. The defects enable the vortex core to successively switch polarization, hindering the transverse domain wall formation. As a result the intrinsic pinning (adiabatic case) as well as the Walker breakdown (non-adiabatic case) are absent. Furthermore, at low currents the domain wall is extrinsically pinned. The successive vortex core switches give rise to a motion of the vortex core in the central region of the nanowire, as also present in perfect nanowires with a degree of non-adiabaticity equal to the Gilbert damping. This explains the consistently lower values $\beta\approx\alpha$ found in experiments based on measurements of the average velocity of vortex domain walls in nanowires.\\ 

These results show that realistic material defects have a significant influence on the domain wall mobility and not only in the creep regime. Therefore, defects should be properly considered when evaluating experimental data and when new concepts are introduced to enhance the domain wall mobility, e.g. by  Spin-hall and Rashba effects.\\

\section{ACKNOWLEDGEMENTS} 
This work is supported by the Flanders Research Foundation (B.V.d.W. and A.V.), the Academy of Finland through an Academy Research Fellowship (L.L., project no. 268302) and through the Centres of Excellence Program (L.L., project no. 251748), Progetto Premiale MIUR-INRIM ``Nanotecnologie per la metrologia elettromagnetica'' (G.D.), and MIUR-PRIN 2010-11 Project2010ECA8P3 ``DyNanoMag'' (G.D.)

\end{document}